\documentclass[aps,twocolumn,showpacs,english]{revtex4-1}
\usepackage{lmodern}
\usepackage[T1]{fontenc}
\usepackage[latin9]{inputenc}
\usepackage{textcomp}
\usepackage{amsmath}
\usepackage{graphicx}
\usepackage{SIunits}
\usepackage{color}

\usepackage{babel}
\usepackage[unicode=true,pdfusetitle,
 bookmarks=true,bookmarksnumbered=false,bookmarksopen=false,
 breaklinks=true,pdfborder={0 0 1},backref=false,colorlinks=false]
 {hyperref}

\begin{document}

\preprint{APS/123-QED}

\title{Static strain tuning of quantum dots embedded in a photonic wire}

\author{   D. Tumanov$^{1,2}$, N. Vaish$^{1,2}$, H.A. Nguyen$^{1,2}$, Y. Cur\'{e}$^{3}$,
  J.-M. G\'{e}rard$^{3}$, J. Claudon$^{3}$, F. Donatini$^{1}$,
    and J.-Ph. Poizat$^{1,2,*}$}

\affiliation{$^1$ Univ. Grenoble Alpes, CNRS, Grenoble INP, Institut NEEL, F-38000 Grenoble, France \\
$^2$ Univ. Grenoble Alpes, CNRS, Grenoble INP, Institut NEEL, "Nanophysique et semiconducteurs" group, 38000 Grenoble, France \\
$^3$ Univ. Grenoble Alpes, CEA, INAC, PHELIQS, "Nanophysique et semiconducteurs" group, F-38000 Grenoble, France \\
* Corresponding author :  jean-philippe.poizat@neel.cnrs.fr}

\date{\today}

\begin{abstract}
We use strain to statically tune the semiconductor band gap of individual InAs quantum dots (QDs) embedded in a GaAs photonic wire featuring very efficient single photon collection efficiency. Thanks to the geometry of the structure, we are able to shift the QD excitonic transition by more than 20 meV by using nano-manipulators to apply the stress. Moreover, owing to the strong transverse strain gradient generated in the structure, we can relatively tune two QDs located in the wire waveguide and bring them in resonance, opening the way to the observation of collective effects such as superradiance.
\end{abstract}

\maketitle


Epitaxial semiconductor quantum dots (QDs) embedded in nanophotonic structures are  very efficient single photon sources (see \cite{Munsch13,Ding16,Somaschi16} and \cite{Senellart17} for a review). However their use in quantum information protocols involving more than two sources has been hindered by the dispersion in energy of different QDs. This dispersion is due to their intrinsically random self-assembly fabrication process \cite{Marzin}, so that two QDs are never alike. QD energy tuning can be achieved using temperature \cite{Gold14}, electric field \cite{Finley,Patel,Bennett}, or material strain \cite{Seidl,Flagg,Wu,Kremer,Trotta15PRL}.
 Temperature tuning is  limited to fine tuning. Electrical control is also suitable for fine tuning and can reach shifts up to $25$ meV \cite{Bennett}.
 Strain tuning can be used for fine tuning \cite{Seidl,Flagg,Kremer,Trotta15PRL} (see \cite{Trotta_Rastelli_review} for a review) and, as temperature and electrical tuning, can enable two-photon interferences with two different QDs \cite{Gold14,Patel,Flagg}. Interestingly, it offers the additional possibility to generate ultra-large shifts up to $500$ meV as demonstrated by Wu et al \cite{Wu} using a diamond anvil cell.
 Such cells are however limited to bulk systems and are unsuitable to QDs embedded in photonic environments.
 Fine strain tuning is usually realized by bonding the bulk QD structure on piezoelectric actuators \cite{Seidl,Flagg,Wu,Trotta15PRL,Trotta_Rastelli_review}, imposing limitations on the structure geometry. Remarkably, Kremer et al \cite{Kremer} managed to achieve up to $1.2$ meV strain tuning for a QD embedded in a nanowire antenna using this bonding technique.



In this paper, we demonstrate  large static strain tuning (up to $25$ meV) of QDs embedded in a photonic waveguide allowing efficient light extraction \cite{Munsch13}. Four years ago, these photonic structures were used by some of us as mechanical oscillator to demonstrate strain-mediated optomechanical coupling \cite{Yeo}. In the present work, the strain is produced \emph{statically} using a nanomanipulator enabling the realization of bright and broadly tunable quantum light sources.
In addition, since the generated strain field features a very large gradient across the wire diameter \cite{deAssis}, our method allows us to bring in resonance two QDs contained in a waveguide and opens interesting perspective for the observation of collective spontaneous emission effects such as superradiance \cite{Goban}.

\begin{figure}[t]
\includegraphics[width=0.8\linewidth]{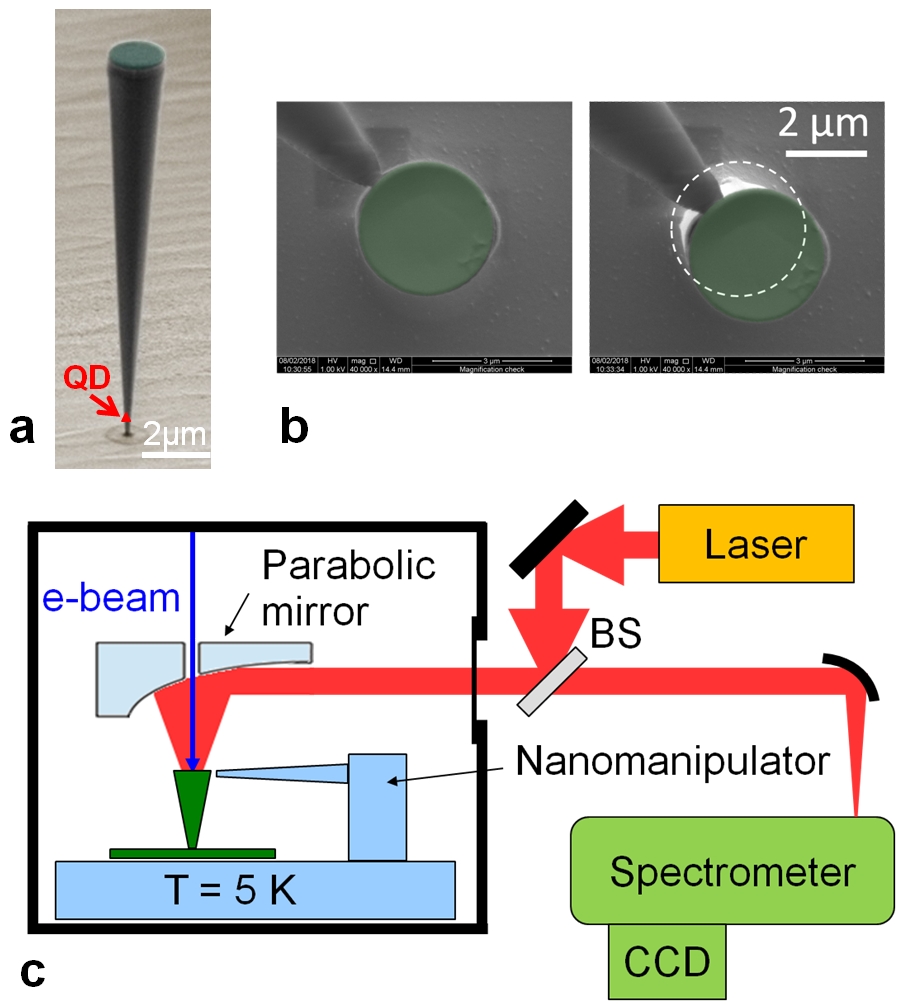}
\caption{ a) Scanning electron microscope (SEM) image of the GaAs photonic waveguide from sample S1. The QDs (red triangle) are located $110$ nm above the basis.  The green top is a false color representing the top facet antireflection coating.  b) SEM top view of  the top facet of a waveguide and the tip of the nanomanipulator. Between the left and right image, the tip has moved and pushed the top of the waveguide by $1\;\mu$m. c) Experimental set-up based on a modified CL set-up for which we have added an excitation laser to enable PL while monitoring the position of the tip with respect to the top facet.}
\label{fig:1}
\end{figure}

\begin{figure}[t]
\includegraphics[width=0.8\linewidth]{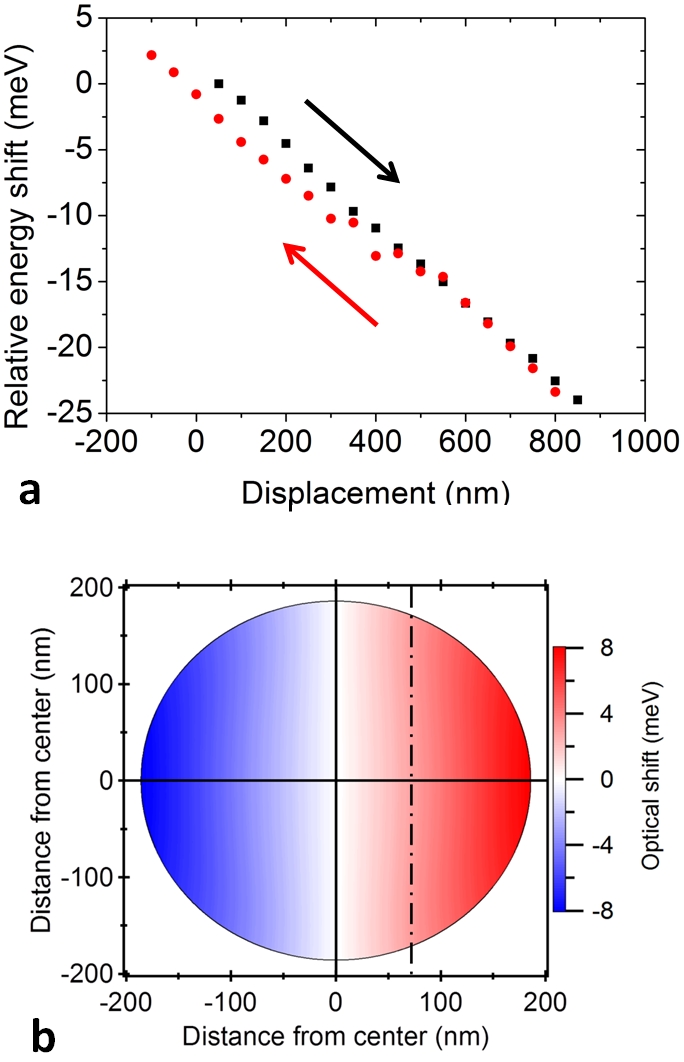}
\caption{a) Relative energy shift as a function of the displacement of nanomanipulator tip obtained from a QD embedded in the waveguide  shown in Fig.\ref{fig:1}a. The black square are for increasing displacement, and the red circles correspond to the way back.  The slight kink on the way back is attributed to the slippage of the tip. b)  Map of the calculated energy shift caused by the strain gradient across the QD plane for a top facet displacement of 100 nm along the horizontal axis. The slope of $31\;\mu$eV/nm of a) corresponds to a QD located on the dash-dotted line at $70$ nm from the zero strain line.  }
\label{fig:Large_shift}
\end{figure}

Our system is based on the GaAs waveguide shown in Fig.\ref{fig:1}a and described in detail in \cite{Munsch13}. The waveguide from sample S1 that we have used for the first experiment (see Fig.\ref{fig:Large_shift}) is $12\;\mu$m high. The top diameter has been measured to $1.69\;\mu$m, and the bottom diameter to $350$ nm. The InAs QDs (around $10$) are located at random positions in a plane at $110$ nm above the basis of the structure that embeds a gold mirror to redirect all the light to the top.
This system has been designed for efficient collection of the single photons emitted by the QDs \cite{Munsch13} as well as the efficient optical addressing of the QD \cite{GNL}. This inverted cone geometry also  features a very strong  optomechanical coupling in which  the mechanical displacement of the top facet induces a large strain in the lower part of the structure affecting the energy of the fundamental excitonic QD transition \cite{Yeo}. In \cite{Yeo}, the top facet motion was caused by the mechanical oscillation of the fundamental flexural mode. In the present work, the displacement is induced statically by a tungsten tip mounted on a nanomanipulator pushing the structure top (see Fig.\ref{fig:1}b,c).  We then excite the QD photoluminescence (PL) to determine the effect of this static stress on the excitonic line energy.

To perform these experiments, we have adapted a cathodoluminescence (CL) set-up to benefit from a scanning electron microscope (SEM) operating at a temperature of $T=5$ K and equipped with a window for optical access (see Fig.\ref{fig:1}c). This allowed us to monitor via SEM the relative position of the nanomanipulator tip with respect to the waveguide (see Fig.\ref{fig:1}b) while using the optical access to carry out PL experiments at a temperature of  $T=5$ K.
The light is collected via an aluminum parabolic mirror, with the object under investigation placed at the focal point of the parabola and centered on the SEM image. This alignment is performed in CL mode (luminescence excitation by the e-beam).
PL implementation in this CL set-up requires the use of a beam splitter between the parabolic mirror and the focalization optics to allow for laser excitation. The exciting laser beam is shaped and aligned to be mode-matched with the CL beam.


The closed-loop nanomanipulator  features a stick and slip coarse mode  that we use for the approach, and a fine mode using an analog piezoelectric scanner with a resolution of about $1$ nm and an amplitude of $1\;\mu$m that we use for pushing on the structure. The nanomanipulator tip is electrically grounded and thermalized using wires connected to the cold plate at a temperature $T=5$ K. The SEM depth of focus allows us to control the tip height within a few hundreds of nm accuracy so that the tip contact point is a few hundreds of nanometers below the top facet. Note that the PL spectrum of the QDs is significantly modified when the tip enters into contact with the waveguide,
 likely because of the rearrangement of the electrostatic environment.

Once the tip is in contact with the waveguide,  the PL spectra is recorded as a function of the position of the nanomanipulator tip as given by the calibrated  closed loop analog piezoelectric driver. Depending on its location within the waveguide, each QD experiences a different spectral strain-induced shift, owing to the strain gradient existing in the QD plane as shown in Fig.\ref{fig:Large_shift}b.
For QDs located far from the zero stress line the spectral shift can be as large as $25$ meV for a top facet displacement of $0.8\; \mu$m as shown in  Fig.\ref{fig:Large_shift}a. Note that we are only limited by the scanning range of the nanomanipulator and that we could probably push further before the waveguide breaks.

A finite element commercial software calculation shows that the main stress component is along the vertical $z$ direction and computes its magnitude for a given top facet displacement. The energy shift $\Delta E$ for a heavy hole exciton in InAs is then given by  the expression \cite{Yu}:
\begin{equation}
\Delta E=\sigma_z \left[ a(S_{11} + 2 S_{12})+b(S_{11}-S_{12}) \right] ,
\end{equation}
with $a= -9$ eV, $b=-2$ eV, $S_{11}=1.14 \times10^{-5}$ MPa$^{-1}$, and $S_{12}=-0.35 \times10^{-5}$ MPa$^{-1}$. For a top facet displacement of $0.8\;\mu$m, our experimental result corresponds to a QD located at a distance of $70$ nm from the zero strain line, undergoing a $420$ MPa stress leading to $\Delta E= 25$ meV.

\begin{figure}[t]
\includegraphics[width=0.7\linewidth]{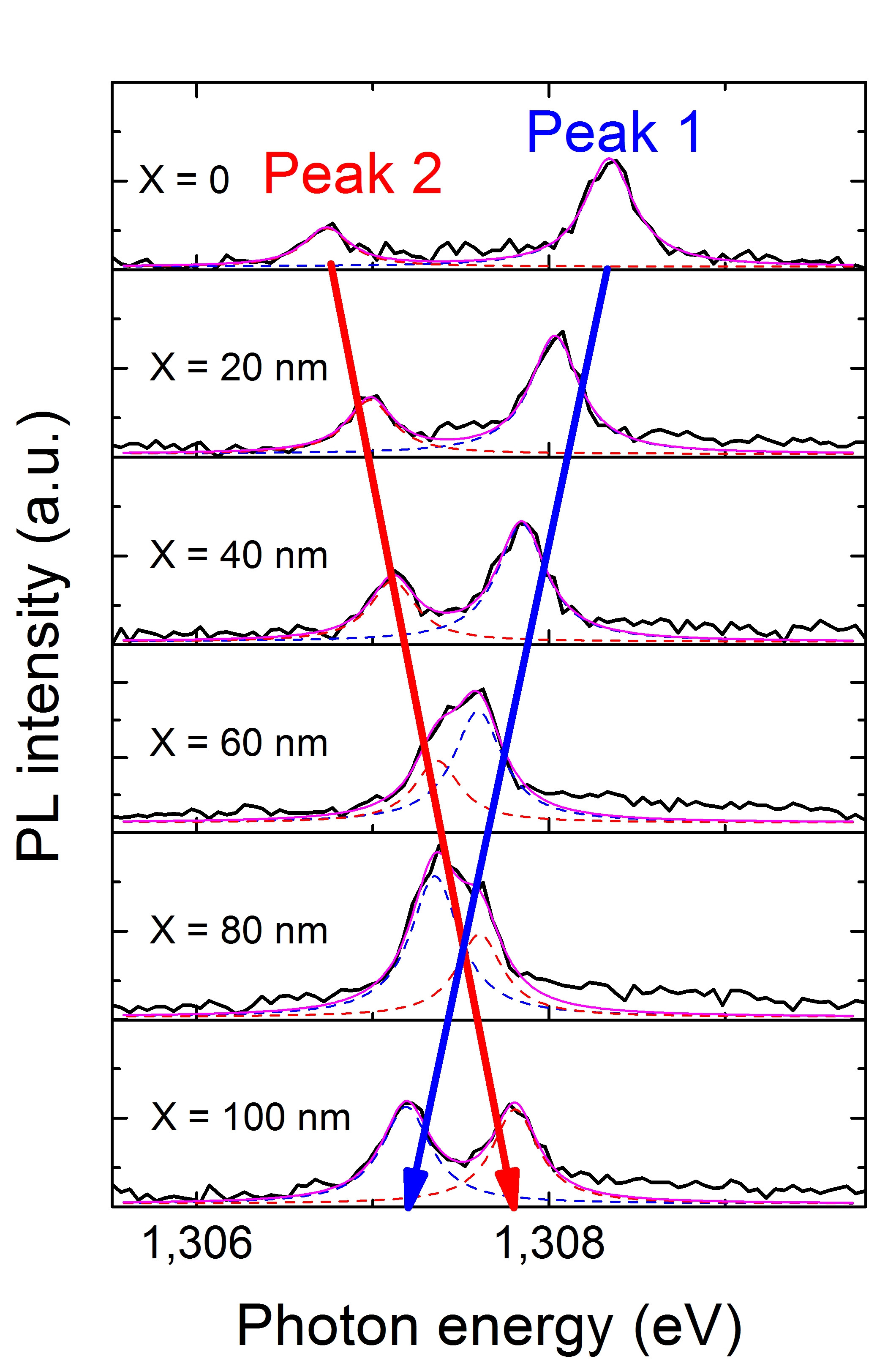}
\caption{PL spectra for different nanomanipulator positions ranging from $X=0$ nm to $X=100$ nm on a different sample  (sample S2) containing more QDs. Sample S2 has already been described in \cite{Yeo} and features a slightly different geometry.
The two peaks correspond to two QDs that are located on either side of the neutral strain line, and therefore undergoing energy shifts of different signs. The two QDs are brought in resonance for $X\simeq 70$ nm. }
\label{fig:merging}
\end{figure}

Since the energy shift depends on the position of the QD within the wire, two QDs located on either sides of the zero-stress  line  undergo shifts of different signs. This can be seen in Fig.\ref{fig:merging} where two QDs are brought in resonance thanks to the static stress applied by the nanomanipulator.  The accuracy of this tuning can be better than the QD life time-limited linewidth of $0.5\;\mu$eV by choosing a QD close to the central neutral line to reduce the sensitivity to the nanomanipulator position and by using a state of the art piezo scanner with position noise as low  $0.1$ nm.
We have also been able to use two independent nanomanipulator tips pushing in two different arbitrary directions. This allowed us to choose and tune the direction of the applied stress. This makes it possible to bring three QDs on resonance in the same waveguide.

Having two or more QDs in resonance in a one-dimensional single mode waveguide  opens up interesting perspective  to investigate collective effects in light-matter coupling. If $g$ is the coupling of a two level system  to a single optical mode, the constructive coherent coupling of  $N$ such two-level systems  to a single optical mode gives rise  to an enhanced light-matter coupling  of $g\sqrt{N}$ of this collective state leading to a  spontaneous emission rate accelerated by a factor $N$ \cite{Goban}. This property can be beneficial for implementing
 quantum memories as mentioned in the landmark paper by Duan et al \cite{DLCZ}.

In summary, we have achieved a large static strain tuning of up to  $25$ meV for a semi-conductor QD embedded in an optically engineered environment using  an original set-up based on photonic wires stressed by nanomanipulators. Thanks to the strong gradient generated  within the structure, we have demonstrated  differential tuning of two QDs coupled to the same single mode waveguide.

\section*{ACKNOWLEDGMENTS}
 Sample fabrication was carried out in the "Upstream Nanofabrication Facility" (PTA) and CEA LETI MINATEC/DOPT clean rooms. We thank Nitin S. Malik for his contribution to the sample fabrication.
   D.T. was supported by a PhD scholarship from the Rh\^{o}ne-Alpes Region,  N.V by a PhD scholarship from Fondation Nanosciences, and H.A.N.  by a PhD scholarship from Vietnamese government. This work was supported by ANR project QDOT  (ANR-16-CE09-0010-01).

\end{document}